\documentstyle[12pt]{article}
\begin{document}

\title{A mathematical model for Neanderthal extinction}
\author{J. C. Flores}
\date{Universidad de Tarapac\'a\\
Departamento de F\'{\i}sica \\
Casilla 7-D Arica\\
Chile\\
(To appears in Journal Of Theoretical Biology, London)}
\maketitle

\baselineskip=18pt

A simple mathematical homogeneous model of competition is used to describe
Neanderthal extinction in Europe. It considers two interacting species,
Neanderthals and Early Modern Men, in the same ecological niche. Using
paleontological data we claim that the parameter of similarity, between both
species, fluctuates between $0.992$ and $0.997$. An extension of the model
including migration (diffusion) is also discussed nevertheless, extinction
of Neanderthal seems unavoidable. Numerical analysis of travelling wave
solutions (fronts) confirms the extinction. The wave-front-velocity is
estimated from linear analysis and numerical simulations confirm this
estimation. We conjecture a mathematical formulation for the principle of
exclusion between competitive interacting species (Gause).

$$
$$

\newpage

Mathematical models for interaction and coexistence between species, yield
non-linear equations which contain a number of rate constants. These
constants are usually determined empirically in controlled experiments, or
by observations in natural environments [1-5].

$$
$$

In this work we consider a system with competitive exclusion. We study the
competition between Neanderthals and men with early modern structure. So,
explicitly, we assume that Neanderthals did not evolve into Early Modern Men.

$$
$$

Neanderthals were very stable in Europe for more than 60.000 years
nevertheless, 40.000 years before our epoch, they were replaced by Early
Modern Men. This massive extinction was completed between 5.000 and 10.000
years depending on the region. It is not clear how modern men appear in
Europe, but some evidence of coexistence in the Levant (see for instance [6]
and [7]), suggests a diffusive process from that region. Moreover, whether
the interaction was direct (war) or not (competition) remains unclear
nevertheless, we assume the last possibility. So, we have two species
competing for the same ecological niche composed of limited nutrient and
territory. The principle of exclusion (Gause) [5,8] can be applied and then
both species cannot coexist.

$$
$$

The following simplified non-linear mathematical model could describe this
biological interaction. Let $N$ be the number of individual, at time $t$,
with Neanderthal characters, and $C$ the one related to Early Modern Men.
Consider the pair of time-evolution equations 
\begin{equation}
\frac{\partial N}{\partial t}=N(F-\beta ), 
\end{equation}
\begin{equation}
\frac{\partial C}{\partial t}=C(F-s\beta ). 
\end{equation}
In this way, we assume the same rate of birth $F(N,C)$ for both species, and
a tiny difference in the constant rate of mortality $\beta $. Namely, we
assume that the parameter of similarity $s$ varies between $0<s\leq 1$,
where $s=1$ means complete similarity. Thus, species $C$ is better adapted
to survive. The limited nutrient reservoir and territory is modeled by the
dependence of $F$ on the variables $N$ and $C$ (see for instance[1]) : 
\begin{equation}
F=\alpha -\delta (N+C) 
\end{equation}
where the growth rate $\alpha $ and the interaction parameter $\delta $ are
positive numbers, and we assume $\alpha >\beta >0$.

$$
$$

Equations (1-3) define a Malthusian-birth-death process with interaction,
and obviously is a crude model. For instance, it does not consider the
diffusion process due to the alleged migration of modern hominid. Moreover,
one might expect the rate constant $\alpha ,\beta $ and $\delta $ to be
affected by changes of ecological nature (temperature, humidity, random
fluctuations, amount of resources, and others). In fact, usually in
population dynamics one deals with systems subjected to random environment
[2,3] where, sometimes, the impact may be drastic (noise-induced
transition). However, equations (1-3) are in accord with the principle of
exclusion. Predictions related to the degree of similarity $s$, between both
species, will be conjectured.

$$
$$

The set (1-3) can be solved partially. The usual linear stability analysis
tell us that the point $(N,C)=(0,0)$ is a unstable node; $(N,C)=(\frac{%
\alpha -\beta }\delta ,0)$ is a saddle point (i.e. unstable); and finally $%
(N,C)=(0,\frac{\alpha -s\beta }\delta )$ is a stable node. All this in
accord with Gause's principle (figure 1). Moreover, a direct integration
gives : 
\begin{equation}
\frac NC=A_0e^{-\beta (1-s)t} 
\end{equation}
where $A_0$ is a constant of integration. As long as $0<s<1$ (already
assumed) the species $N$ disappear, and the time of extinction $\tau $ can
be related to the parameters by 
\begin{equation}
\tau =\frac 1{\beta (1-s)}. 
\end{equation}
Using the above relationship, paleontological data for the extinction time $%
\tau $ (i.e. $5000<\tau <10000$, years), and the life-time for individual ($%
30<1/\beta <40$, years), then $s$ fluctuates between $0.992<s<0,997$. It is
instructive to compare this result, for instance, with the parameter of
similarity between man and chimpanzee [9] where $s^{\prime }\sim 0.975$.
This last parameter is related to the sequence of nucleotides in DNA, and
then not necessarily connected to $s$. On the other hand, following
Reef.[10], Neanderthal industry (silex-knife) did require about 111
percussions (4 stages) against 251 (9 stages) for Cro-Magnon. So the
comparison of the number of percussion (for stage) gives $s^{\prime \prime }=%
\frac{111}4/\frac{251}9$ $\sim 0.995$. A number curiously close of our
parameter $s$. 
$$
{} 
$$

Turning to the two species model (1-3), a more realistic case requires
migration. This can be carried-up by adding a diffusive term onto (2), and
neglecting the mobility of $N$ in a first approach. In a more quantitative
form, for instance, consider a migration term like to this one considered
originally by Volterra [1,2], i.e. adding up a positive constant $m$ onto
(2) :

$$
\frac{\partial C}{\partial t}=C(F-s\beta )+m,\quad \quad \quad \quad \quad
(2^{\prime }) 
$$
the linear analysis of (1,2') and (3) shows that the only stable point is $%
(N,C)\sim (0,\frac{\alpha -S\beta }\delta +\frac m\alpha )$, corresponding
to the extinction for the species $N$. Nevertheless, a modification on the
time of extinction $\tau $ (5) is expected in this case. So, a constant
migration term does not stop Neanderthal extinction.%
$$
{} 
$$

In a more realistic approach, we can consider a diffusive term like to $%
D\partial _{xx}C$ added to (2) and look for travelling wavefront solution.
Namely we consider solutions like to $N(x-vt)$ and $C(x-vt)$ in the
population variables where $v$ is the velocity of propagation. In this case
using the variable $z=x-vt$ , the evolution equations become,

\begin{equation}
-v\frac{\partial N}{\partial z}=N\left( \alpha -\beta -\delta \left(
N+C\right) \right) , 
\end{equation}

\begin{equation}
-v\frac{\partial C}{\partial z}=C\left( \alpha -s\beta -\delta \left(
N+C\right) \right) +D\frac{\partial ^2C}{\partial z^2}. 
\end{equation}

The linear stability analysis tell us that the point $(N,C)=(\frac{\alpha
-\beta }\delta ,0)$ is a saddle point with one stable manifold.
Nevertheless, the condition

\begin{equation}
v^2>4D\left( 1-s\right) \beta 
\end{equation}
is necessary because any physical solution requires $C\geq 0$. On the other
hand the point $(N,C)=(0,\frac{\alpha -s\beta }\delta )$ is a unstable node
and, finally, the point $(N,C)=(0,0)$ is always stable (figure 2). For this
last point, the condition

\begin{equation}
v^2>4D\left( \alpha -s\beta \right) 
\end{equation}
must be imposed ($N,C\geq 0$). Moreover, remark the invariance under
velocity inversion $(v\rightarrow -v)$ and coordinate inversion $%
(z\rightarrow -z)$ in (6,7). So, for any solution with velocity $v$ we can
found one other with velocity $-v$.

$$
{} 
$$

From the above discussion, there is the possibility of travelling wave
solutions connecting the point $(\frac{\alpha -\beta }\delta ,0)$ and $(0,%
\frac{\alpha -s\beta }\delta )$. Numerical calculations confirm this
possibility. Figure 3 shows two front-expansion for species $C$ and the
backward motion for $N$ is superposed. The stability of this wavefront
solution was tested numerically using different extended initials conditions
for $N$. So, numerical solutions confirm the extinction. Remark, like to the
Fisher equation [5], we can use the inequality (9), more stronger than (8),
to obtain a first estimation of the wavefront velocity, i.e. $v\sim \sqrt{%
4D\left( \alpha -s\beta \right) }$. Numerical calculations confirm this
estimation. Figure 4 shows the variation of the velocity wave front for
different values of $\sqrt{D}$. It was carried up assuming $\int
C(x,t)dx\propto vt$ after the transient producing the stable front (figure
3).

$$
{} 
$$

Finally, the deterministic model (1-3) with interaction is simple and
similar to those proposed in [1]. Nevertheless, as long as we assume the
validity of the exclusion principle (Gause), other deterministic models
[2-4] cannot gives us very different results. Namely, the exclusion
principle guaranties the extinction of species $N$. In fact, from equations
(1-3) we have the inequality

\begin{equation}
\frac 1C\frac{\partial C}{\partial t}>\frac 1N\frac{\partial N}{\partial t} 
\end{equation}
which tell us that no-equilibrium points exist like to $(N_0,C_0)$, where $%
N_0\neq 0$ and $C_0\neq 0$. Moreover, the above inequality (10) guaranties
the instability of equilibrium points like to $(N_0,0)$. In this way, the
above inequality can be interpreted as a mathematical formulation of the
exclusion principle, and then we can expect its validity for other
deterministic models describing the interaction between $N$ and $C$.

$$
$$

$$
{} 
$$
ACKNOWLEDGMENTS. This work was partially supported by grant FONDECYT 394 000
4.

\newpage

FIGURE CAPTIONS 
$$
$$
$$
$$

Figure 1: A numerical sketch of competition between Neanderthal ($N$) and
Early Modern Men ($C$). A simple mathematical model is described by
equations (1-3). Species $C$ fills gradually the ecological niche of $N$
(arbitrary unities).

$$
$$

Figure 2: A linear stability analysis of equation (6-7) with a diffusive
term. $(N,C)=(\frac{\alpha -\beta }\delta ,0)$ is a saddle point and $%
(N,C)=(0,\frac{\alpha -s\beta }\delta )$ is a unstable node. In this way,
there is a possible solution connecting these points. The point $(0,0)$ is
stable.

$$
{} 
$$

Figure 3: A typical numerical simulation of the solution of (1-3) with a
diffusive term $D\partial _x^2C$ add to (2). The central pick corresponds to
the expanding front for $C$. The other two, are related to Neanderthal
backward motion ($0<X<1000$). The stability of these wave-fronts have been
tested numerically using different extended initial conditions for $N$
(arbitrary unities).

$$
{} 
$$

Figure 4: A numerical calculation of the wave front velocity $v$ for
different diffusion coefficients ($\sqrt{D}$). The assumption, to obtain
this graphic, is the number of total individual $C$ grows like to $vt$ after
the transient (i.e. $t\rightarrow \infty )$. This calculation confirms our
estimation for the wave front velocity.

\end{document}